\documentclass[preprint,12pt,3p]{elsarticle}

\usepackage{amssymb}
\usepackage{lineno,hyperref}
\modulolinenumbers[5]
\usepackage{graphicx}
\usepackage{dcolumn}
\usepackage{bm}
\usepackage{amsmath}
\usepackage{color}

\usepackage{numcompress}

\journal{Applied Acoustics}

\begin{document}

\begin{frontmatter}

\title{On the wave dispersion and non-reciprocal power flow in space-time traveling acoustic metamaterials}

\author[mysecondaryaddress]{M. A. Attarzadeh}

\author[mysecondaryaddress]{H. Al Ba'ba'a}

\author[mysecondaryaddress]{M. Nouh\corref{mycorrespondingauthor}}
\cortext[mycorrespondingauthor]{Corresponding author}
\ead{mnouh@buffalo.edu}

\address[mysecondaryaddress]{Dept. of Mechanical \& Aerospace Engineering, Univeristy at Buffalo (SUNY), Buffalo, NY}

\begin{abstract}
This note analytically investigates non-reciprocal wave dispersion in locally resonant acoustic metamaterials. Dispersion relations associated with space-time varying modulations of inertial and stiffness parameters of the base material and the resonant components are derived. It is shown that the resultant dispersion bias onsets intriguing features culminating in a break-up of both acoustic and optic propagation modes and one-way local resonance band gaps. The derived band structures are validated using the full transient displacement response of a finite metamaterial. A mathematical framework is presented to characterize power flow in the modulated acoustic metamaterials to quantify energy transmission patterns associated with the non-reciprocal response. Since local resonance band gaps are size-independent and frequency tunable, the outcome enables the synthesis of a new class of sub-wavelength low-frequency one-way wave guides.  
\end{abstract}


\end{frontmatter}


\section{Introduction}
\begin{figure}[h]
\centering
\includegraphics[width=0.6\linewidth]{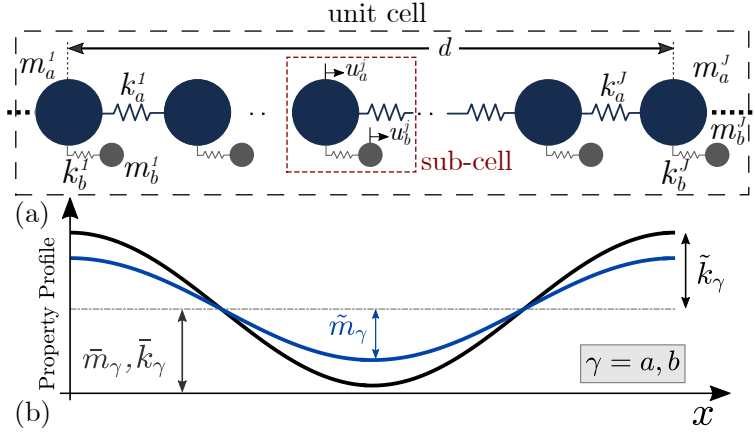}
\caption{\label{fig:NR_mm} (a) Lumped realization of a locally resonant acoustic metamaterial (LRAM) and (b) mass and stiffness modulation profile within a unit cell (bottom)}
\end{figure}

The last few decades have witnessed a spurt of activity investigating the use of metamaterials to realize unique solutions to problems in vibroacoustic mitigation, wave cloaking, focusing, guidance, and others \cite{pai_huang_book, deymier2013acoustic, Hussein2014}. Locally resonant acoustic metamaterials (LRAMs) are sub-wavelength structures that exhibit mechanically tunable, size-independent, low frequency band gaps \cite{liu_sonic}. In their common form, LRAMs comprise a base (outer) structure that houses a series of uniformly distributed inner resonators (Fig.~\ref{fig:NR_mm}a), which contribute to the rise of unique dispersion properties. Local resonance band gaps in LRAMs stem from their ability to significantly attenuate incident excitations over a narrow frequency spectrum at the vicinity of the resonators' eigenfrequencies \cite{huang2009negative}. As such, LRAMs have been recently investigated in the context of discrete lumped mass systems \cite{huang2011study, Huang2010}, elastic bars \cite{Pai2010, xiao2012longitudinal}, flexural beams \cite{sun2010, Nouh2014, xiao2013flexural, Zhu2014}, as well as 2D membranes and plates \cite{towards, wang2013, Nouh2015}.

\vspace{0.2cm}

Owing to their periodic nature, band structures of LRAMs can be computed using a Bloch-Floquet wave solution. These structures convey the wave dispersion relations $\omega(\mu)$, where $\omega$ is the frequency and $\mu$ is the dimensionless wavenumber. Due to elastodynamic reciprocity, band structures of LRAMs are symmetric about $\mu=0$ implying that waves travel from point $A$ to $B$ in the same manner they would travel from $B$ to $A$ \cite{NR_elasto}. Breaking this reciprocity in 1D systems creates a bias in the band structures intended to force waves to travel differently in opposing directions \cite{zanjani2014one,swinteck2015bulk}. Non-reciprocity in metamaterials have been very recently utilized to synthesize, among others, acoustic guides \cite{NR_acoustics2} and static displacement amplifiers \cite{static_NR}. Means to induce non-reciprocal behavior include introduction of large nonlinearities, topological features, and material fields that travel in time and space \cite{popa_NR, trainiti2016non}. The latter has been recently demonstrated in elastic metamaterials using a perturbation approach \cite{nassar2017elastic}. Although very challenging, several efforts have recently investigated achieving material variations in time using negative capacitance piezoelectric shunting \cite{ruzzene2017time}, inductance-based resonance control \cite{cardella2016manipulating}, and magnetoelastic materials \cite{ansari2017applicationNew}. In this work, we build on the work developed in \cite{Vila2017363} for non-resonant space-time traveling phononic lattices to develop a mathematical framework that captures and predicts non-reciprocal dispersion physics in lumped time-traveling LRAMs. After analytically deriving the asymmetric wave dispersion relations based on a defined unit-cell, we validate the framework using the finite band structures reconstructed from the actual response of an LRAM chain of a known length. Furthermore, we present a structural intensity analysis of the non-reciprocal LRAMs and derive the power flow maps associated with the non-reciprocal energy transmission in the LRAM as a result of the imposed modulations.

\vspace{0.2cm}

This note is organized in four sections. Following the introduction, we begin by deriving the governing equations for spatiotemporally modulated mass and stiffness properties for both the base and the resonant components of a lumped LRAM to obtain the non-reciprocal dispersion relations. Through numerical simulations, a 2-Dimensional Fourier Transform (2D-FT) is then performed to validate the obtained band structure derived analytically. To further investigate the non-reciprocal behavior, in the subsequent section we investigate the LRAM using the energy-based structural intensity analysis (SIA) to capture the power flow patterns within the non-reciprocal range. Finally, the conclusions are briefly summarized.

\section{Dispersion Relations}
\subsection{Mathematical Formulation}
To onset acoustic non-reciprocity, the parameters of the LRAM have to undergo a traveling-wave like modulation. As such, we begin by deriving mass $m$ and stiffness $k$ properties which travel simultaneously in time and space. Contrary to the conventional unit cell definition, we define a unit cell of a subset of lumped masses, spanning a length $d$, that constitute a single cycle of property variation (Fig.~\ref{fig:NR_mm}a). We also denote each spring-mass system and its resonator as a sub-cell. Consequently, we consider harmonic variation of $m$ and $k$ as follows

\begin{equation}
m_{\gamma}^j(t)=\bar{m}_{\gamma}+{{{\tilde{m}}}_{\gamma}}\cos \left( {{\omega }_{0}}t+{{\kappa }_{0}}j \right) \label{eq:mass_variation}
\end{equation}
\begin{equation}
k_\gamma^{j} (t) =\bar{k}_\gamma+ \tilde{k}_\gamma \cos (\omega_0 t+\kappa_0 j) \label{eq:N4}
\end{equation}

\noindent where, as depicted in Fig.~\ref{fig:NR_mm}b,  $j = 1, \dots , J$ is the sub-cell index and $J$ is the total number of sub-cells within a unit cell. Also $\gamma = a,b$ refers to the base masses and local resonators, respectively. $\bar{k}_\gamma$ and $\bar{m}_\gamma$ are the average values of both variations while $\tilde{k}_\gamma$ and $\tilde{m}_\gamma$ are the oscillatory components. Further, $\omega_{0}$ and $\kappa _{0}={2\pi}/{J}$ represent the temporal and spatial modulation frequencies. In practice, such modulations can be physically realized via piezoelectric or magnetoelastic actuation \cite{ansari2017analyzing}. Equations governing the motion of the $j^{th}$ sub-cell can be derived as
\begin{equation}
m_a^j\ddot{u}_a^j+(k_a^j+k_a^{j+1}) u_a^j-k_{a}^j u_a^{j-1}-k_a^{j+1} u_a^{j+1}+k_b^j (u_a^j-u_b^j)=0 \label{eq:BM_EOM}
\end{equation}
\begin{equation}
m_b^j \ddot{u}_b^j+k_b^j(u_b^j-u_a^j)=0 \label{eq:LR_EOM}
\end{equation}

\noindent where $u_a^j$ and $u_b^j$ denote the base mass and resonator displacements, respectively. Using the Floquet-Bloch theorem \cite{Bloch1929,slater1958interaction}, and exploiting the LRAM's periodicity, the unit cell displacement can be related to its adjacent ones via $u_{\gamma}^{J+1}=u_{\gamma}^{1}{e}^{-i\mu}$ and $ u_{\gamma}^{0}=u_{\gamma}^{J}{{e}^{i\mu}}$, where $i$ is the imaginary unit. Upon establishing periodic boundary conditions, the motion equations of the entire cell can be represented in compact matrix notation as
\begin{equation} 
\mathbf{M}(t) \ddot{\mathbf{u}} + \mathbf{K}(t,\mu) \mathbf{u} = \mathbf{0}
\label{eq:Reduced_EOM}
\end{equation}

\noindent where $\mathbf{u}=\{u_a^1, u_a^2, \dots, u_a^J|u_b^1, u_b^2, \dots, u_b^J\}^{T}$ is the displacement field, and $\mathbf{K}(t,\mu)$  and $\mathbf{M}(t)$ are the unit cell stiffness and mass matrices. Being periodic functions of time, both $\mathbf{K}$  and $\mathbf{M}$ can be expanded using a complex Fourier series as follows
\begin{equation} \label{eq:M_expansion}
\mathbf{M}(t) = \sum_{p=-\infty }^{\infty }\mathbf{\hat{M}}_{p} e^{ip{{\omega }_{0}}t}
\end{equation}

\begin{equation} \label{eq:K_expansion}
\mathbf{K}(t,\mu) = \sum\limits_{p=-\infty }^{\infty}\mathbf{\hat{K}}_{p} e^{ip{{\omega }_{0}}t}
\end{equation}

\noindent where $\mathbf{\hat{M}}_{p}$ and $\mathbf{\hat{K}}_{p}$ are the corresponding Fourier matrix coefficients. Henceforth, we assume a harmonic solution with a time-modulated amplitude of the following form
\begin{equation}
\mathbf{u}={e}^{i\omega t} \sum_{n=-\infty}^{\infty} \mathbf{\hat{u}}_n {{e}^{in{{\omega }_{0}}t}}
\label{eq:u_expansion}
\end{equation}

\noindent where $\mathbf{\hat{u}}_n$ is the $n^{th}$ vector of displacement amplitudes. Substituting Eqs.~(\ref{eq:M_expansion}), (\ref{eq:K_expansion}), and (\ref{eq:u_expansion}) into (\ref{eq:Reduced_EOM}), and employing harmonic function orthogonality, we obtain
\begin{equation} \label{eq:infty_quad_EVP}
\sum\limits_{n=-N }^{N} \bigg(\mathbf{A}_{s,n}^{(2)} \omega^2+\mathbf{A}_{s,n}^{(1)} \omega + \mathbf{A}_{s,n}^{(0)} \bigg) \mathbf{\hat{u}}_n =\mathbf{0}
\end{equation}

\noindent in which $N$ is the truncated limit of the infinite series and $s$ is an arbitrary integer within the interval $[-N,N]$. $\mathbf{A}_{s,n}^{(q)}$, with $q=0,1,2$, is a $2J \times 2J$ matrix for any $s$ and $n$ combination. It is defined as
\begin{equation} \label{eq:A_0}
\mathbf{A}_{s,n}^{(0)}= n^2 \omega_0^2 \mathbf{\hat{M}}_{s-n} - \mathbf{\hat{K}}_{s-n}
\end{equation}
\begin{equation} \label{eq:A_1}
{\mathbf{A}_{s,n}^{(1)}}=2 n \omega_0\mathbf{\hat{M}}_{s-n}
\end{equation}
\begin{equation} \label{eq:A_2}
{\mathbf{A}_{s,n}^{(2)}}=\mathbf{\hat{M}}_{s-n}
\end{equation}

\noindent Eqs.~(\ref{eq:infty_quad_EVP}) through (\ref{eq:A_2}) can be combined into a quadratic eigenvalue problem \cite{Nouh2015smart}
\begin{equation} \label{eq:quad_EVP}
\bigg(\mathbf{\Phi_{2}} \omega^{2} + \mathbf{\Phi_{1}} \omega + \mathbf{\Phi_{0}} \bigg) \mathbf{\hat{U}} = \mathbf{0}
\end{equation}

\noindent where the new vector $\hat{\mathbf{U}}$ is obtained by stacking all $\mathbf{\hat{u}}_n$ for $n=-N $ to $N$ sequentially. The block matrix $\mathbf{\Phi_{q}}$ is of size $2J(2N+1)\times 2J(2N+1)$ and each of its elements is a sub-matrix defined by
\begin{align} \label{eq:N17}
\mathbf{\Phi_{q}} (s,n)={\mathbf{A}_{s,n}^{(q)}} && q=0,1,2
\end{align}

\noindent Eq.~(\ref{eq:quad_EVP}) requires the matrix multiplied by $\hat{\mathbf{U}}$ to be singular in order to yield a non-trivial solution, which describes the acoustic wave dispersion in the LRAM lattice. If an index $p$ is defined such that $p = s-n$, explicit forms of $\mathbf{\hat{M}}_p$ and $\mathbf{\hat{K}}_p$, which constitute ${\mathbf{A}_{s,n}^{(q)}}$, can be found as
\begin{equation}
\mathbf{\hat{M}}_p=
\begin{bmatrix}
\mathbf{M}^p_a & \mathbf{0} \\ \mathbf{0} & \mathbf{M}^p_b
\end{bmatrix}
\label{eq:Complete_EOM_1}
\end{equation}

\begin{equation}
\mathbf{\hat{K}}_p=
\begin{bmatrix}
\mathbf{K}^p_a+\mathbf{K}^p_b & \mathbf{-K}_b^p \\ \mathbf{-K}^p_b & \mathbf{K}^p_b
\end{bmatrix}
\label{eq:Complete_EOM_2}
\end{equation}

\noindent such that
\begin{equation}
\mathbf{\hat{M}}^{p}_\gamma
=\bigg(\bar{m}_\gamma \delta_{p,0}+\frac{\tilde{m}_\gamma}{2} (\delta _{p,-1} e^{i \kappa_0 j}+\delta_{p,1}e^{-i \kappa_0j})\bigg) \odot \mathbf{I}
\end{equation}
\begin{equation}
\mathbf{\hat{K}}^{p}_b
=\bigg(\bar{k}_b\delta_{p,0}+\frac{\tilde{k}_b}{2} (\delta _{p,-1} e^{i \kappa_0 j}+\delta_{p,1}e^{-i \kappa_0 j})\bigg) \odot \mathbf{I}
\end{equation}
\begin{equation}
\mathbf{\hat{K}}^{p}_a
= \bar{k}_a\delta_{p,0} \mathbf{\Psi}_0+\frac{\tilde{k}_a}{2} (\delta _{p,-1} \mathbf{\Psi}_{-1}+\delta_{p,1}\mathbf{\Psi}_{+1})
\end{equation}

\noindent where $ \mathbf{I}$ is the identity matrix and $\odot$ denotes element-wise multiplication. The definitions of $\mathbf{\Psi}_{+1}$, $\mathbf{\Psi}_{-1}$ and $\mathbf{\Psi}_{0}$ are given by
\begin{align}
\mathbf{\Psi}_{\ell}=
\begin{bmatrix}
\psi^\ell_1+\psi^\ell_2  & -\psi^\ell_2  &  & & &  -\psi^\ell_1 e^{i \mu} \\
-\psi^\ell_2 & \psi^\ell_2+\psi^\ell_3 & -\psi^\ell_3 & & & \\
& \ddots  & \ddots &  \ddots &  &    \\
& &  &  &  &  \\
& &  &  & &  -\psi^\ell_{J}    \\
-\psi^\ell_1 e^{-i \mu}& &  & & -\psi^\ell_{J} & \psi^\ell_{J}+\psi^\ell_1\\
\end{bmatrix} 
\label{eq:Psi_1_-1}
\end{align}

\noindent where $\psi_j ^ \ell = e^{-i \ell \kappa_0 j}$ and $\ell = -1,0,1$. 

\vspace{0.2cm}

\subsection{Numerical Validation}
\begin{figure}[h]
\centering
\includegraphics[width=0.94\linewidth]{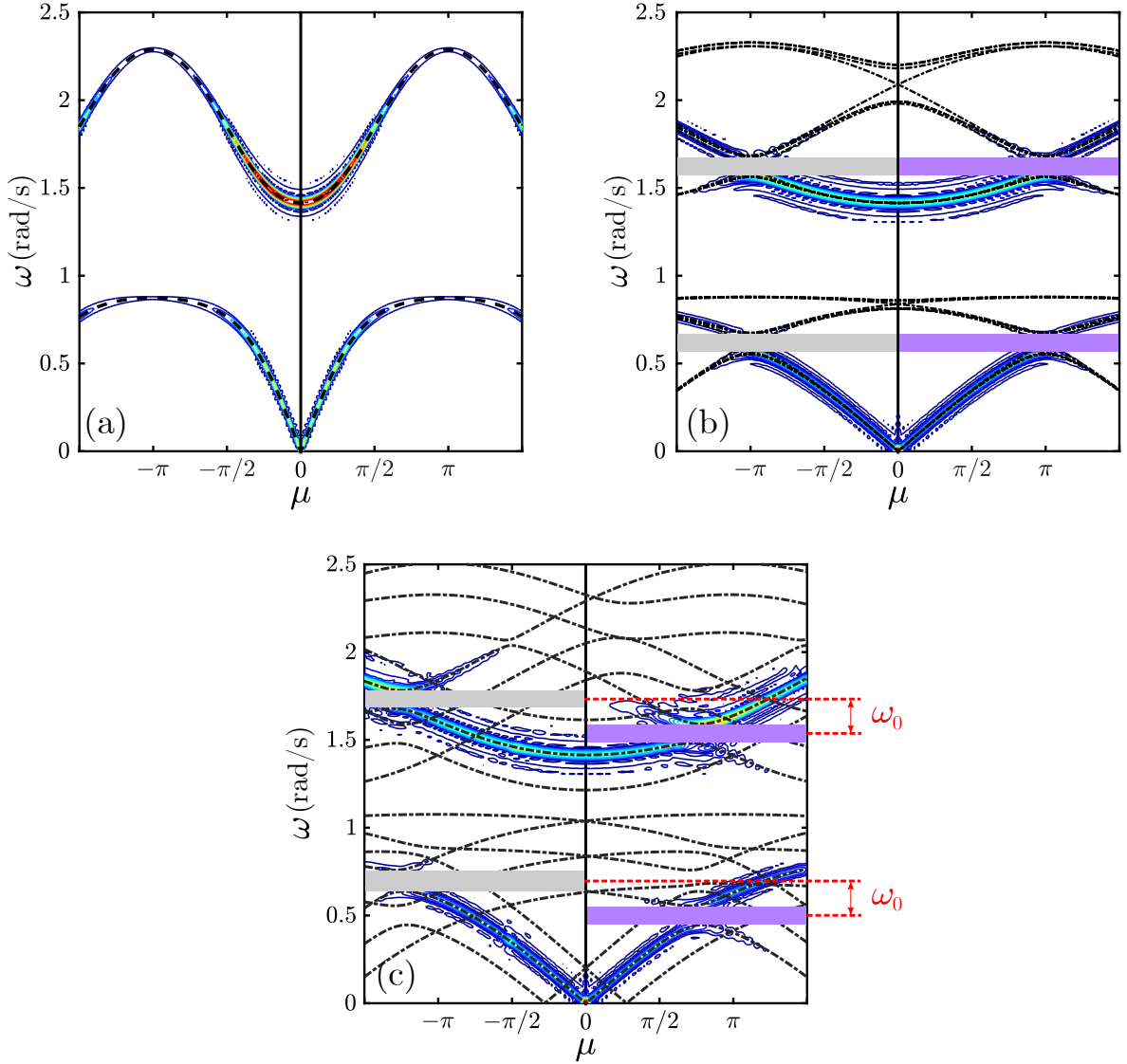}
\caption{\label{fig:Dispersion} Dispersion bands from Eq.~(\ref{eq:quad_EVP}) (black lines) and by using a 2D-FT of the time-domain response (contours). Shaded regions indicate forward (purple) and backward (grey) band gap ranges. Results shown are for (a) An unmodulated LRAM with $J=1$. (b) Spatially modulated LRAM with $\omega_0=0$, $J=3$. (c) Spatio-temporally modulated LRAM metamaterial with $\omega_0=0.2$, $J=3$.}
\end{figure}

To validate the derived dispersion relations, we performed a transient numerical simulation of a spatiotemporally modulated LRAM using the following parameters: $\bar{m}_a = \bar{m}_b = \bar{k}_a = \bar{k}_b =1 $, $\tilde{m}_a = \tilde{m}_b = \tilde{k}_a = \tilde{k}_b = 0.3$, $J=3$ and $\omega_0 = 0.2$. The analysis is carried out on a sufficiently large lattice consisting of $305$ sub-cells. The LRAM is excited precisely at its mid-span to avoid anomalies associated with lattice asymmetry that can influence results. Dispersion contours constructed from the response of the finite LRAM can be evaluated using a discrete variant of the conventional 2D-FT \cite{airoldi2011design} given by
\begin{equation} \label{eq:2DFT} 
{U(\mu,\omega)}=\frac{d}{J} \sum_{(j)}^{} e^{i\mu x_j} \int \limits_{-\infty}^{\infty} e^{-i\omega t} u(j,t)dt
\end{equation}

\noindent where $x_j$ is the position of each sub-cell with respect to the global coordinate system. The approach uses the numerical response of an LRAM subject to a broadband transient wave-packet to excite the entire frequency range of interest. During which, the resultant displacement field is recorded for windowed time and space. The results are given in Fig.~\ref{fig:Dispersion} for an unmodulated (i.e. $\tilde{k_a} = \tilde{k_b} = \tilde{m_a} = \tilde{m_b} = 0$), spatially modulated (i.e. $\omega_0=0$), and a spatiotemporally modulated LRAM, and are compared to band structures computed from the derived dispersion relations. The unmodulated case (with $J=1$) serves as a benchmark and shows a traditional band structure of a locally resonant acoustic metamaterial \cite{huang2009negative}. The presence of the resonators splits the structure into two dispersive branches, an acoustic (lower) branch and an optic (upper) branch depicting the two possible oscillation modes, with a band gap (approximately spanning $0.874<\omega<1.414$) in between.

\newpage

\begin{figure}[h]
\centering
\includegraphics[width=\linewidth]{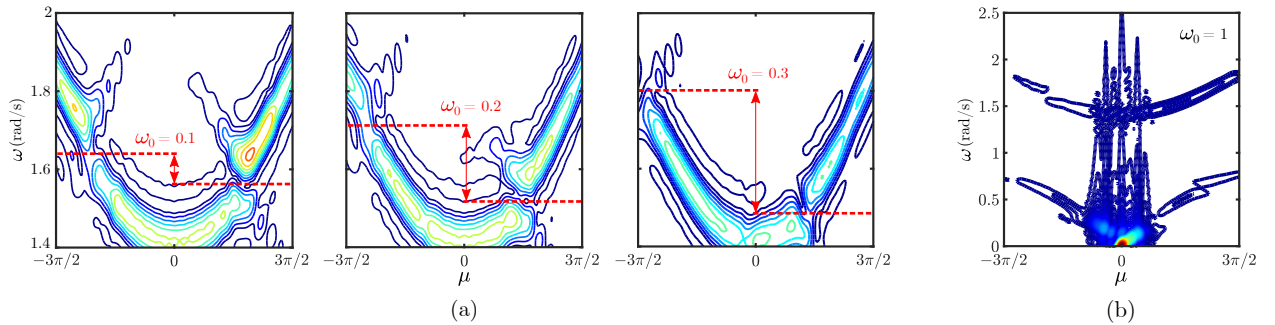}
\caption{\label{fig:modulation_freq} (a) Optic dispersion branch of the spatiotemporally modulated LRAM at $\omega_0=0.1$, 0.2, and 0.3 (arrows mark the ranges of the emergent non-reciprocal band gaps). (b) Full dispersion diagram at $\omega_0=1$}
\end{figure}

Upon applying a spatial modulation to the LRAM parameters, two new band gaps (shaded areas in Fig.~\ref{fig:Dispersion}b) emerge by breaking-up both the acoustic and optic branches. Acoustic reciprocity is, however, fully maintained and the newly emerging band gaps spread across the entire wavenumber space. With temporal modulation added, the width (frequency range) of the band gaps remains almost unchanged but they shear and eventually separate at the wavenumber origin $\mu=0$. The split-up band gaps yield one way propagation structures at their respective frequencies. Depending on the latter, the LRAM can act as a forward- or backward-only diode for incident excitations. It is worth noting that not all the analytically computed dispersion branches appear in the 2D-FT contours. This has been similarly reported in modulated phononic crystals where an eigenvector-based weighting factor was used to filter out redundant branches \cite{Vila2017363}. The choice of the modulation frequency depends on the desired magnitude of disparity between the forward and the backward band gaps. A gradual increase in $\omega_0$ results in a larger separation of the non-reciprocal band gaps as shown in the optic branch displayed in Figure \ref{fig:modulation_freq}a for $\omega_0=0.1$, 0.2, and 0.3. While the frequency shift between the two opened band gaps remains equal to $\omega_0$, a decrease in the forward band gap frequency and an increase in the backward band gap frequency take place. As a consequence, the split of the dispersion branch which onsets the non-reciprocal behavior shifts to a slightly larger wavelength for the forward mode and to a smaller wavelength for the backward mode as evident in the same figure. With that in mind, it is necessary to avoid large increases in the modulation frequency $\omega_0$ which risk the possibility of discontinuities and time-growing waves that are associated with unstable interactions between propagating waves and the imposed modulations in time-space periodic media as outlined in \cite{cassedy1963dispersion, cassedy1967dispersion}. Such effects can render the effort to establish an observable non-reciprocal behavior futile. This is also shown in Figure \ref{fig:modulation_freq}b when the modulation frequency matches the local resonance of the unmodulated system.

\vspace{0.3cm}

\begin{figure}[h!]
\centering
\includegraphics[width=0.7\linewidth]{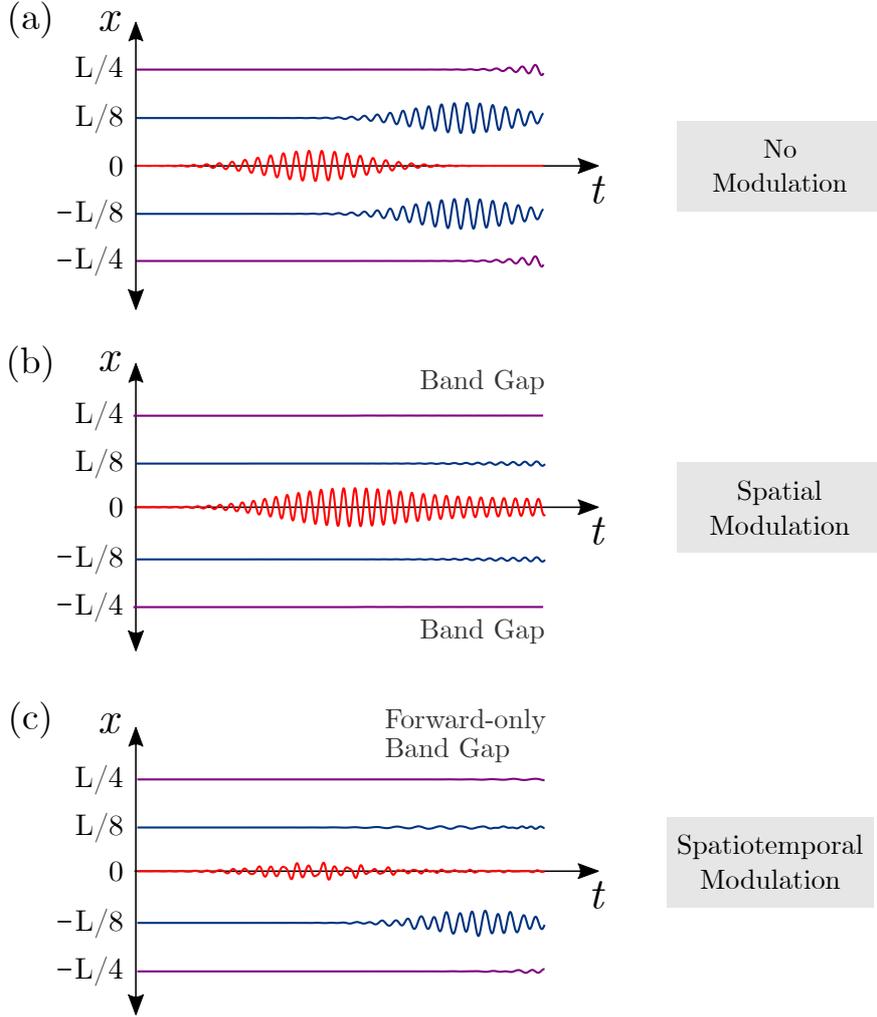}
\caption{\label{fig:Time_response} Time-response of the LRAM's displacement field at different spatial locations: (a) An unmodulated LRAM, (b) Spatially modulated LRAM excited at $\omega_c = 0.58$ and (c) Spatiotemporally modulated LRAM excited at $\omega_c = 0.5$}
\vspace{-2pt}
\end{figure}

Fig.~\ref{fig:Time_response} shows all the preceding features via the time-response of the LRAM's displacement field at different spatial locations along its length $L$. The LRAMs here are excited at their mid-point with a narrow band excitation centered around a frequency $\omega_c$ to illustrate forward and backward propagating modes.

\section{Power Flow Analysis}
To further illustrate the non-reciprocity of the LRAM, a structural intensity analysis (SIA) is exercised to provide visual snapshots of power flow in the modulated lattice at different time instants. In the $x$-direction, the complex transmitted power $P(t)$ is calculated as a function of the internal forces $N_x(t)$ and the velocity vector $\dot{u}_x(t)$ using \cite{pavic1, cho2016structural}
\begin{equation}
P(t) = - N_x(t) \dot{u}_x(t)
\end{equation}

\noindent and has been recently reported as an effective measure of energy attenuation in local resonance band gaps of LRAMs \cite{AlBabaa2016a, albabaa2017}. The SIA is carried out on three cases of the considered LRAM: Case 1: An LRAM with no modulation. Cases 2 and 3: A spatiotemporally modulated LRAM with the same modulation parameters used to generate the dispersion bands. Similar to Fig.~\ref{fig:Time_response}, the LRAMs are excited at the mid-span with a narrow band excitation centered around a frequency $\omega_c$. In the second case, $\omega_c$ is outside the one-way band gaps observed in Fig.~\ref{fig:Dispersion}, while it lies within it in the third case.

\begin{figure}[h]
\centering
\includegraphics[width=\linewidth]{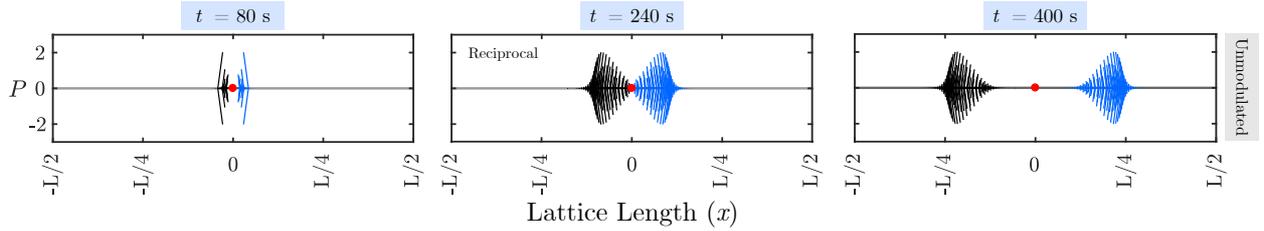}
\caption{\label{fig:PF1} Transient power flow in the unmodulated LRAMs at 80, 240, and 400 seconds (Excitation frequency is $\omega_c=0.5$ and circular dots mark excitation location)}
\end{figure}

\begin{figure}[h]
\centering
\includegraphics[width=\linewidth]{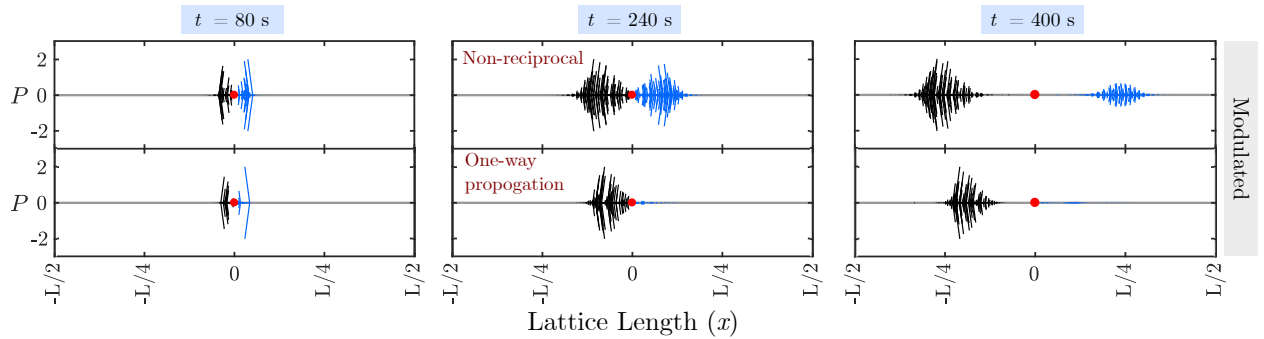}
\caption{\label{fig:PF2} Transient power flow in the modulated LRAMs at 80, 240, and 400 seconds. Upper row: Spatiotemporally modulated and excited at $\omega_c=0.25$. Bottom row: Spatiotemporally modulated and excited at $\omega_c=0.5$ (circular dots mark excitation location)}
\end{figure}

The computed power flow diagrams are shown in Figs.~\ref{fig:PF1} and \ref{fig:PF2} using a \textit{quiver} representation in which the power flow variation along the length of the LRAM is given as arrows that capture the direction and magnitude (arrowhead height) of $P$. Excitation source is marked by a red circular dot at the lattice's mid-point and the variations are plotted at three different time instants. The diagrams demonstrate the flow of the energy packet (emerging from the excitation location) as it travels along the length of the LRAM lattice in both directions. In the unmodulated case (Fig.~\ref{fig:PF1}), the packet travels symmetrically in both directions about the LRAM's origin demonstrating, as expected, full acoustic reciprocity. At all three time instants shown, $P(x)$ and $P(-x)$ are identical at any arbitrary location along the length $L$. Further, the excitation applied (at $\omega_c=0.5$) does not coincide with the local resonance band gap, as can be verified from Fig.~\ref{fig:Dispersion}a, and as a result power flows freely to both ends of the LRAM. In the spatiotemporally modulated LRAMs, cases 2 and 3, non-reciprocal power flow is evidently observed. In the upper row of Fig.~\ref{fig:PF2}, $P$ flows asymmetrically about the origin and attenuates faster on the forward propagation side. Due to the space-time parameter modulations, the energy packet emerging from the excitation at $x=0$ effectively sees two different mediums on both sides at each time instant of the transient simulation. Finally, in case 3, the central frequency of the exciting force ({$\omega_c=0.5$}) hits a band gap in the lower (acoustic) positive dispersion branch and consequently exhibits a forward-only band gap which is clearly picked up by the power flow diagrams displayed in the bottom row of Fig.~\ref{fig:PF2}.

\section{Conclusions}
In conclusion, the analytical dispersion relations have been derived for a locally resonant acoustic metamaterial (LRAM) exhibiting traveling-wave-like modulations of both the mass and stiffness properties. The dispersion curves were used to characterize the emergent non-reciprocal dispersion patterns in the media of such systems. It was shown that the dispersion bias as a result of the parameter modulation can onset intriguing features in the LRAM bands, including loss of parity and asymmetric dispersion about the wave vector origin, as well as break-up of both the acoustic (lower) and optic (upper) dispersion branches yielding forward- (or backward-) only local resonance band gaps. The presented mathematical framework was verified using dispersion contours constructed from the transient response of finite LRAMs under different excitations using a Fourier-transform based approach. Furthermore, using a structural intensity analysis (SIA), the non-reciprocal behavior was validated and the power flow diagrams in the modulated time-traveling LRAMs were shown to capture the evolution of non-reciprocal energy transmission patterns. Since LRAMs are sub-wavelength systems, such findings can be used to instigate a new class of low-frequency structures that can yield non-reciprocal properties at moderate size scales.

\section*{Acknowledgments}

This work was partially supported by the US National Science Foundation under CMMI award 1647744.

\bibliographystyle{elsarticle-num}

\bibliography{sample}

\end{document}